\documentclass[a4paper]{article}

\usepackage{INTERSPEECH2020}
\usepackage{url}
\usepackage{xcolor}
\usepackage{cite}
\usepackage{numprint}
\npthousandsep{,}

\title{UIAI System for Short-Duration Speaker Verification Challenge 2020}
\name{Md Sahidullah$^1$, Achintya Kumar Sarkar$^2$, Ville Vestman$^3$, Xuechen Liu$^{1,3}$, Romain Serizel$^1$, Tomi Kinnunen$^3$, Zheng-Hua Tan$^4$, Emmanuel Vincent$^1$}
\address{
  $^1$Universit\'{e} de Lorraine, CNRS, Inria, LORIA, F-54000, Nancy, France\\
  $^2$Indian Institute of Information Technology, Sri City, India\\
  $^3$School of Computing, University of Eastern Finland, Joensuu, Finland\\
  $^4$Aalborg University, Aalborg, Denmark}
\email{md.sahidullah@inria.fr}

\begin{document}

\maketitle
\begin{abstract}
In this work, we present the system description of the UIAI entry for the short-duration speaker verification (SdSV) challenge 2020. Our focus is on Task 1 dedicated to text-dependent speaker verification. We investigate different feature extraction and modeling approaches for automatic speaker verification (ASV) and utterance verification (UV). We have also studied different fusion strategies for combining UV and ASV modules. Our primary submission to the challenge is the fusion of seven subsystems which yields a normalized minimum detection cost function (minDCF) of 0.072 and an equal error rate (EER) of 2.14\% on the evaluation set. The single system consisting of a pass-phrase identification based model with phone-discriminative bottleneck features gives a normalized minDCF of 0.118 and achieves 19\% relative improvement over the state-of-the-art challenge baseline. 
\end{abstract}


\section{Introduction}
\emph{Automatic speaker verification} (ASV) is the task of verifying whether a speech utterance has been spoken by a claimed speaker \cite{kinnunen2010overview,hansen2015speaker}. State-of-the-art ASV systems show promising performance when several minutes of audio data have been collected in controlled conditions for both enrollment and verification. The amount of enrollment and verification speech is an important factor \cite{poddar2017speaker}. While having more speech typically improves recognition performance, short-duration utterances are often preferable for practical deployment. The \emph{short-duration speaker verification} (SdSV) challenge 2020 primarily focuses on the duration factor where speakers are enrolled and verified with a few seconds of audio data \cite{SdSVplan}. Besides, the speech corpus for the challenge was recorded in realistic environments, and the collection protocol was designed to incorporate various kinds of noises in the speech corpus which introduced mismatches between the enrollment and the verification phases. The challenge has two independent tasks. Our entry focuses on Task 1, which concerns \emph{text-dependent} ASV. 

In text-dependent ASV, the spoken phonetic contents for enrollment and verification are assumed to be identical. Typically, a short sentence or phrase is shared by all users. However, considering the practical fact that the test speaker can also utter a wrong phrase \cite{larcher2014text,lee2015reddots}, there may be four types of trials, respectively defined as \emph{target correct} (TC) where the target speaker utters the correct phrase, \emph{target wrong} (TW) where the target speaker utters a different phrase, \emph{impostor correct} (IC) where an impostor utters the same phrase as in speaker enrollment, and \emph{impostor wrong} (IW) where an impostor utters a different phrase compared to speaker enrollment. For evaluation purposes, TC trials are considered as genuine trials to be \emph{accepted}, and the remaining three as impostor trials to be \emph{rejected}.

Although methods developed for text-independent ASV are applicable to text-dependent ASV without any modification, they do not generalize well \cite{larcher2014text}. In particular, the performance severely degrades in the TW condition due to the similarities in speaker information. The solutions proposed for text-dependent ASV verify the speaker identity and the spoken phrase in an integrated manner. Here, the phrase information is incorporated by capturing contextual information with a \emph{hidden Markov model} (HMM) \cite{larcher2014text,sarkar2016text,zeinali2017hmm,zeinali2016deep,dey2017template}, \emph{dynamic time warping} (DTW) \cite{dey2017template,he2016low} and pass-phrase identification \cite{Sarkar2017}. Alternatively, standalone UV and ASV modules can be developed, and fused together at the score or decision level \cite{Kinnunen2016}.

We investigate both strategies for this challenge. Our single system applies joint spoken phrase and speaker identity verification where phrase information is incorporated with a \emph{phrase-dependent background model} (PBM) \cite{Sarkar2017}. The primary system integrates modules developed for separate tasks. Here, we adopt a \emph{cascade fusion} strategy where UV is performed before ASV. We first compute the decision threshold associated with the \emph{equal error rate} (EER) for UV on the development set. A verification trial is assigned with an arbitrarily low ASV score if its UV score is lower than the threshold. On the other hand, a trial is passed to the ASV module for scoring if its UV score is greater than or equal to the threshold. The threshold computed on the development set is adopted for scoring on the evaluation set. Our UIAI team is a multi-site collaboration involving four research labs. Given the emphasis on text-dependent ASV, we investigate different \emph{utterance verification} (UV) and text-dependent approaches. We develop an \emph{i-vector}-based UV method that includes channel variability compensation. We introduce a text-dependent ASV method employing \emph{phone-based bottleneck features} with PBM. We also explore how a standard ASV system can be improved in short-duration conditions with different frame-level \emph{acoustic features} and utterance-level \emph{speaker embeddings}. Finally, we study different system combination strategies suitable for combining UV and ASV systems. Our \emph{primary system}, which is a multi-level fusion of \emph{seven} different subsystems, has achieved the \emph{fifth} rank in the challenge whereas the \emph{single system} has shown substantial improvement over the two challenge baselines.

The rest of the paper is organized as follows. Section \ref{Section:Sysdesc} describes different subsystems developed for the challenge. Section \ref{Section:ExperimentalSetup} describes the experimental setup. Section \ref{Section:Results} presents the experimental results. Finally, we conclude in Section \ref{Section:Conclusion}.

\section{System description}\label{Section:Sysdesc}

In this section, we summarize the subsystems used for our primary submission to the SdSV challenge. 

\subsection{Utterance verification}
Our UV system relies on \emph{speaker embeddings}. Although speaker embeddings mainly encode speaker information, they contain a considerable amount of information about the spoken content \cite{ProbingRajASRU,wang2017does}. This makes them potentially useful for UV tasks (besides ASV). The UV task in the SdSV challenge is a closed-set task as there are no out-of-set phrases. We use an i-vector representation \cite{dehak2010front} and a \emph{probabilistic linear discriminant analysis} (PLDA) back-end for this task. The setup is similar to that commonly used in ASV scenarios, except that we treat utterance identifiers (rather than speakers) as the class labels. The i-vectors are projected onto a $9$-dimensional space using \emph{linear discriminant analysis} (LDA). Whitening and length normalization are applied to the projected i-vectors. We then use Gaussian PLDA with a full-rank subspace to compute the pairwise UV score between the average i-vector of the claimed phrase and the i-vector of the test phrase. Finally, we apply score normalization suitable for this closed-set scenario where the number of possible hypotheses is fixed. We use Max norm which subtracts the maximum score of the other (competing) phrases from the hypothesized phrase score \cite{Kinnunen2016}. We also tried Mean norm but it performed worse. 

\subsection{Speaker verification}
We develop four standalone ASV systems based on \emph{x-vector-PLDA} \cite{snyder2018x} and \emph{Gaussian mixture model-universal background model} (GMM-UBM) \cite{ubmmap} approaches. 

\renewcommand*{\arraystretch}{1.0}
\begin{table}[b!]
\caption{{\it Architecture of speaker embedding extractor network. The layers from 1 to 8 and layer 10 are followed by leaky ReLU activations and batch normalization. The embeddings are extracted from layer 10 before applying the activation function.} \label{table:xvector_architecture}}
\centering
\begin{tabular}{l l c c}
\toprule
\# & Layer type & CNN kernel size & Output dim. \\
\midrule
1 & TDNN-SE & $5$ & $512$ \\
2,3 & TDNN-RES-SE & $5$ & $512$ \\
4,5 & TDNN-RES-SE & $7$ & $512$ \\
6,7 & TDNN-RES-SE & $1$ & $512$ \\
8 & TDNN-SE & $1$ & $128$ \\
9 & LDE aggregation & --- & $\numprint{8192}$ \\
10 & FC & --- & $512$ \\
11 & FC-softmax & --- & \#speakers \\
\bottomrule
\centering
\end{tabular}
\end{table}
\renewcommand*{\arraystretch}{1.0}

\textbf{X-vector-PLDA system}: The network architecture for extracting x-vector speaker embeddings is given in Table~\ref{table:xvector_architecture}. It differs from the original x-vector architecture by adding \emph{squeeze-and-excitation} (SE) \cite{hu2018squeeze} modules to each frame-level \emph{time-delay neural network} (TDNN) layer. In addition, three of the frame-level layers are replaced by \emph{residual} (RES) blocks \cite{he2016deep}. The global mean and standard deviation pooling layer is replaced with a \emph{learnable dictionary encoder} (LDE) \cite{zhang2017deep,cai2018novel}. The LDE layer is similar to a GMM: it assigns features into components and computes statistics locally. We used a variant of LDE with a diagonal covariance matrix shared across all components. For more details of the speaker embedding extractor, see \cite{vestman2020neural}. The trials are scored with a PLDA module \cite{garcia2011analysis} trained on the training embeddings. Finally, adaptive symmetric score normalization (AS-norm) \cite{cumani2011comparison} is applied.

\textbf{GMM-UBM systems}: The success of GMM-UBM systems in speaker verification with short utterances \cite{poddar2017speaker} as well as text-dependent ASV \cite{Kinnunen2016,larcher2014text,sahidullah2016local} motivates us to explore this approach for the SdSV challenge. Our GMM-UBM systems are similar to standard GMM-UBM systems except that the UBMs are trained in a more efficient way. Instead of training a UBM on the full dataset, we train $10$ GMMs on audio data for each of the $10$ phrases independently and create the UBM by merging the components of the $10$ GMMs and normalizing the mixture weights to unit sum. In the following, we consider three GMM-UBM systems relying on different acoustic features.

\subsection{Joint utterance and speaker verification}
We perform joint verification of spoken content and speaker identity based on \emph{pass-phrase dependent background models} (PBMs) \cite{Sarkar2017}. In this approach, PBMs are first derived from the GMM-UBM using maximum a posteriori (MAP) adaptation with phrase-specific audio data. During the enrollment phase, target speaker models are created by MAP adaptation from the best-matched PBM instead of the single UBM in a GMM-UBM system. The best-matched PBM is found in the maximum likelihood (ML) sense. In the test phase, we first determine the best-matched PBM for the test utterance. Finally, we compute the log-likelihood ratio score between the target model and the best-matched PBM. Our primary submission includes two PBM systems based on two different sets of \emph{bottleneck features} (BN) extracted with deep neural networks (DNNs).



\begin{figure}[b!]
\includegraphics[height=2.6cm,width=11.5cm]{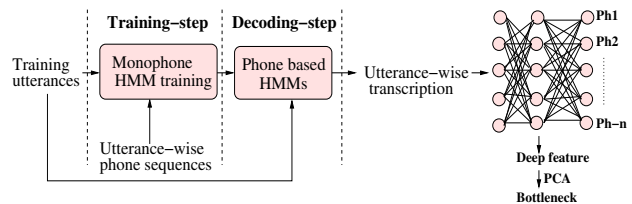}
\caption{Phone-BN feature extraction system.}
\label{fig:HMM_dnn}
\end{figure}

\textbf{Phone-discriminative bottleneck (Phone-BN) features} are extracted by a DNN trained on phone labels obtained using an HMM. We use the in-domain audio data of the SdSV challenge along with phone sequence information. We chose the flat-start method \cite{htkbook} for HMM training as phone-level alignments are not available. First, we train HMM-based \emph{monophone models} by pooling all the utterances and their corresponding phone sequences. The parameters of the phone-based HMMs are then re-estimated in an iterative manner using the \emph{Baum-Welch} algorithm. We consider $42$ HMMs that correspond to the number of unique phones in the training data. We consider a $3$-state (excluding start and end-state) left-to-right topology without skipping state transitions. We also model silences and pauses between words. We generate phone-level alignments and discard silences and pauses. Then we train a DNN to classify phones. The frame-level output of one hidden layer \cite{DBLP:journals/taslp/SarkarTTSG19} is then projected using \emph{principal component analysis} (PCA) to obtain lower-dimensional Phone-BN features. Figure~\ref{fig:HMM_dnn} illustrates the Phone-BN feature extraction system.

\textbf{Stream-wise time-contrastive learning based BN (sTCL-BN) features} are extracted in a similar way, except that the phone classes used as DNN training targets are found in an unsupervised way~\cite{Sarkar2017nipsw,DBLP:journals/taslp/SarkarTTSG19}. First, all the speech utterances are randomly concatenated into a single \emph{stream}. This stream is segmented into fixed-duration chunks that capture short-term context. Given the desired number $N$ of classes, $N$ chunks are taken at a time, and data points in the $n$-th chunk are assigned to class label $n$ where $n \in \{1,2,...,N\}$. The process is repeated until all data points have been assigned. A segment-based clustering algorithm \cite{Sarkar2017nipsw,DBLP:journals/taslp/SarkarTTSG19} is applied to group similar chunks together and update the class labels until convergence. We then train a DNN to discriminate the obtained labels. The frame-level output of a hidden layer is projected into a low dimensional space using PCA to obtain sTCL-BN features.

\section{Experimental setup}\label{Section:ExperimentalSetup}

\subsection{Dataset description}
The audio data for the SdSV challenge was created from the DeepMine dataset collected by crowdsourcing \cite{deepmine2018odyssey,deepmine2019asru}. Task 1 involves speech files from $963$ speakers as \emph{in-domain} audio data. There is no development set which could be used for parameter tuning and optimization. We have created a development set by randomly choosing a subset of $63$ speakers from this in-domain data. Similarly to the evaluation set, we enroll each speaker with three utterances for phrase-specific targets. If adequate data for a speaker-phrase combination is not available, we disregard that target model. The remaining sentences are used for test. We obtain a total of $519$ target models and $\numprint{4810}$ speech utterances for test in the development set. The number of trials is summarized in Table~\ref{tab:example}. Similarly to the evaluation set, there is no cross-lingual trial. We have also discarded same-gender trials by clustering i-vectors for the $63$ speakers (averaged over all utterances) into two \emph{pseudo-gender} classes and keeping the trials for which the speakers in the enrollment model and the test utterance fall into the same class.


\begin{table}[!th]
  \caption{Number of trials per condition in the development set.}
  \label{tab:example}
  \centering
  \renewcommand{\arraystretch}{1.2}
  \begin{footnotesize}
  \begin{tabular}{|c|c|c|c|c|}
    \hline
    \textbf{TC} & \textbf{TW}& \textbf{IC} &\textbf{IW} & \textbf{Total}\\
    \hline
     \numprint{4810}                       &  \numprint{19236} & \numprint{119737} & \numprint{478946} & \numprint{622729}            \\
    \hline
  \end{tabular}
  \end{footnotesize}
\end{table}

The remaining in-domain data from $900$ speakers consisting of $\numprint{94661}$ utterances are considered for system development in addition to the other permitted audio data, such as VoxCeleb and LibriSpeech. The evaluation set consists of $\numprint{12404}$ enrollment models, $\numprint{69542}$ utterances, and $\numprint{8306700}$ trials.

\subsection{Dataset and parameters for system training}
\label{subsec:params}
We refer the seven subsystems in the UIAI entry as S1--S7. 

\textbf{S1} is the UV system and it uses \emph{mel-frequency cepstral coefficient} (MFCC) features. We extract $20$-dimensional static MFCCs, apply a RASTA filter, and compute deltas and double-deltas to create $60$-dimensional features. Utterance-level \emph{cepstral mean and variance normalization} (CMVN) is applied after discarding non-speech frames using energy-based activity detection (SAD). A $512$-component UBM is trained on in-domain training data. The \textbf{T}-matrix is estimated with $600$ factors on the same audio data. The extracted i-vectors are projected to $9$ dimensions using LDA based on phrase labels.

\textbf{S2} is the x-vector-PLDA ASV system. The x-vector extractor is trained on YouTube audio data from VoxCeleb-1 \cite{nagrani2017voxceleb} and VoxCeleb-2 \cite{chung2018voxceleb2}. Recordings from the same YouTube video source are concatenated together. The scoring back-end (PLDA, centering, whitening, AS-norm) is trained on the in-domain data. For PLDA, the training labels are pairs of speaker and phrase IDs. Both in-domain and VoxCeleb data are augmented $5$-fold using Kaldi's \cite{povey2011kaldi} augmentation recipes which include reverberation and additional noise, babble, or music. The input features for the embedding extractor are $60$-dimensional, cepstral mean normalized MFCCs extracted with Kaldi. Kaldi's energy based VAD is applied to remove non-speech frames.

\textbf{S3--5} are the GMM-UBM based ASV systems. Ten phrase-specific $512$-component GMMs are trained on the in-domain data and merged into a $5120$-component UBM. The target speaker models are created by MAP adaptation with a relevance factor of $3$. We use three different acoustic front-ends. System \textbf{S3} is based on $60$-dimensional MFCCs including deltas and double-deltas. System \textbf{S4} uses linear-frequency cepstral coefficients (LFCCs) with the same dimension. System \textbf{S5} uses $66$-dimensional \emph{overlapped block transform coefficients} (OBTCs) computed with two blocks of sizes $9$ and $13$ \cite{sahidullah2012design}. The pre- and post-processing stages are identical for the three feature sets. We use $20$ mel filters and retain the energy coefficients. The features are processed with RASTA filtering and utterance-level CMVN. We do not apply SAD since this degraded performance.

\begin{table*}[h]
  \caption{Results (EER in \% / minDCF) in the development set for the individual subsystems included in the UIAI primary submission.}
  \begin{footnotesize}
  \begin{center}
  \label{Table:ResultsDevelopment}
  \centering
  \renewcommand{\arraystretch}{1.2}
  \begin{tabular}{|c|c|c|c|c|c|c|c|c|}
    \hline
    \textbf{ID} & \textbf{Method}  & \textbf{Short-term features} & \textbf{Task} & \textbf{TW} &\textbf{IC} & \textbf{IW} & \textbf{Pooled} & \textbf{Avg.} \\
    \hline
    \textbf{S1}  &   i-vector     &    MFCC         &  UV   & 0.08 / 0.005 & 51.64 / 1.00 & 0.08 / 0.005 & 18.29 / 1.00 & 17.27 / 0.337\\
    \textbf{S2}  & x-vector     & MFCC                &  ASV   &9.11 / 0.575	& 0.82 / 0.041 &	0.12 / 0.004 & 0.93	/ 0.085	& 3.35 / 0.207\\
    \textbf{S3}  &  GMM-UBM & MFCC                   &  ASV  & 8.35 /	0.431&	0.91 /	0.040 &	0.60 / 	0.019&	1.36 /	0.090&	3.29 /	0.164\\
    \textbf{S4}  & GMM-UBM      & LFCC              &  ASV   &10.69 / 0.535	& 1.27 / 0.057 &	0.77 / 0.028	& 1.68 / 0.116	& 4.24 / 0.206\\
    \textbf{S5}  & GMM-UBM  & OBTC                   &  ASV   &7.82 /	0.405	& 0.85 / 0.035 &	0.60 / 0.016	& 1.22 / 0.085	& 3.09 / 0.152\\
    \textbf{S6}  & PBM &        Phone-BN            &  Both  &0.07 / 0.008	& 1.31 / 0.051 &	0.01 / 0.001	& 0.81 /	0.029 & 0.46 / 0.020\\
    \textbf{S7}  & PBM      &   sTCL-BN            &  Both  &0.14 / 0.011	& 1.74 / 0.062	& 0.01 / 0.001	& 1.07 / 0.038	& 0.63 / 0.025\\
    \hline
  \end{tabular}
  \end{center}
  \end{footnotesize}
\end{table*}

\begin{table*}[h]
  \caption{Results (EER in \% / minDCF) of single and primary systems in the development and evaluation sets. The x-vector based baseline achieves 9.05\% EER and 0.529 minDCF while the i-vector/HMM baseline achieves 3.49\% EER and 0.146 minDCF.}
  \begin{footnotesize}
  \begin{center}
  \label{Table:ResultsSubmission}
  \centering
  \renewcommand{\arraystretch}{1.2}
  \begin{tabular}{|c|c|c|c|c|c||c|}
    \hline
    \textbf{System}   & \textbf{TW} &\textbf{IC} & \textbf{IW} & \textbf{Pooled} & \textbf{Avg.} & \textbf{Eval set. (Pooled)}\\
    \hline
    Single (\textbf{S6})  &0.07 / 0.008	& 1.31 / 0.051 &	0.01 / 0.001	& 0.81 /	0.029 & 0.46 / 0.020 & 3.83 / 0.118\\
    Primary (fusion \textbf{S1-S7}) & 0.06 / 0.006 & 0.32 / 0.015 & 0.00 / 0.000 & 0.17 / 0.007 & 0.13 / 0.007 & 2.14 / 0.072\\
    \hline
  \end{tabular}
  \end{center}
  \end{footnotesize}
\end{table*}

\textbf{S6} is the PBM-based joint verification system with Phone-BN features. The DNN feature extractor consists of $7$ fully-connected layers with $1024$ neurons and \emph{sigmoid} activation. Its inputs are $57$-dimensional RASTA-filtered MFCCs including deltas and double-deltas with a context of $11$ frames. The number of outputs is $42$. The DNN is trained on the in-domain data using CNTK \cite{CNTK}. We compute BN features by projecting the output of the second hidden layer on each time frame into a $57$-dimensional vector using PCA. The PCA matrix is computed on $\numprint{10000}$ randomly selected utterances from the training part of LibriSpeech. We use the open-source \emph{robust voice activity detector} (rVAD) \cite{TanSD20} to discard non-speech frames from the enrollment and test utterances before applying utterance-level CMVN. The HMM used for extracting the phone labels does not discard non-speech frames, but it uses utterance-level CMVN. We used HTK \cite{htkbook} to build the HMM system. 
The Phone-BN features are then used with the PBM system where a gender-independent GMM-UBM with $2048$ Gaussian components is trained using $60,000$ utterances from LibriSpeech. We create the target models using MAP adaptation with $3$ iterations. We adapt only the mean vectors with a relevance factor of $10$. 
 
\textbf{S7} is the PBM-based joint verification system with sTCL features. The DNN feature extractor uses the same input features as \textbf{S6}. We use rVAD \cite{TanSD20} to discard non-speech frames, and we consider $N=10$ unsupervised classes as in our previous study \cite{DBLP:journals/taslp/SarkarTTSG19}. The DNN and the PBMs are trained on the same audio data using the same hyper-parameters as \textbf{S6}.

\subsection{Performance evaluation}
The primary metric for the challenge is the \emph{normalized minimum detection cost function} (minDCF) \cite{SdSVplan}. The EER is also reported. These are computed by treating TC trials as genuine and by pooling the remaining three as impostor. We computed those metrics on the development set using our own scoring script while the challenge organizers computed performance on the evaluation set. We also report the performance for the three sub-conditions and their average values on the development set.

\begin{figure}[ht]
\centering
\includegraphics[scale=0.4]{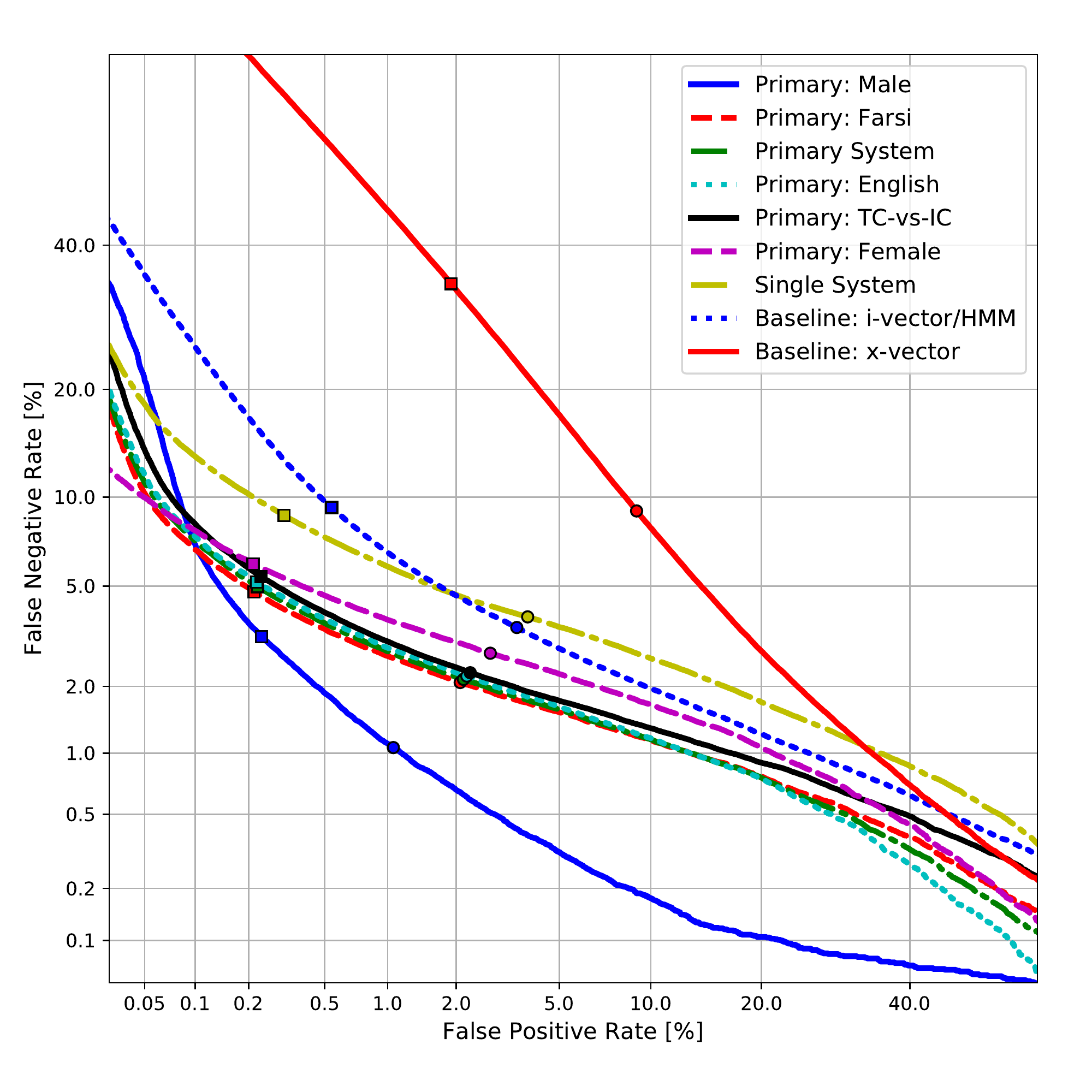}
\caption{DET plots of the UIAI systems along with baselines.}
\label{Fig:detplot}
\end{figure}

\section{Results}\label{Section:Results}

The ASV performance of the individual subsystems on the development set is shown in Table~\ref{Table:ResultsDevelopment}. The first row shows the ASV result of the UV system 
\textbf{S1} which performs well in wrong pass-phrase conditions (TW and IW). However, the performance for the IC condition is random with about $50\%$ EER as this system does not use speaker information. The next row shows the performance for the x-vector-based ASV system \textbf{S2}. Although it yields promising performance in the IC and IW conditions, it performs poorly in the TW condition. Similarly, the GMM-UBM systems perform relatively well in the IC and IW conditions but they fail in the TW condition. Our results also indicate that the GMM-UBM systems give competitive performance compared to the x-vector system. The short-term OBTC features outperform MFCCs in all cases. Out of the two PBM-based methods, the one based on Phone-BN features performs consistently better and both of these methods outperform other subsystems in terms of average EER and minDCF. \textbf{S6} achieves the lowest average and pooled EERs as well as minDCFs. For this reason, we select it as the single system for the challenge. 

We built the primary system submitted to the challenge by combining the modules for UV and ASV. We fuse \textbf{S1}, \textbf{S6} and \textbf{S7} for the UV task and all the subsystems except \textbf{S1} for the ASV task. The subsystems for each task are combined by linear score weighting where the weights are optimized on the development set by linear search. The UV and ASV system are then combined by cascade fusion as described earlier. The trials with wrong pass-phrases as detected by the fused UV system are assigned an ASV score of $-100$, while the trials with correctly detected pass-phrases are retained with fused ASV score.

Table~\ref{Table:ResultsSubmission} summarizes the results achieved by the single system and the primary system on the development and evaluation sets. The single system outperforms the two challenge baselines, especially in terms of minDCF. Figure~\ref{Fig:detplot} shows the DET plots of the submitted systems along with the baselines.

\section{Conclusions}\label{Section:Conclusion}
We have described the UIAI systems submitted to the SdSV challenge 2020 for text-dependent ASV. The systems developed are a fusion of different subsystems using various front-ends and back-ends. To deal with the pass-phrase verification problem, we combine the UV system with ASV in a cascade mode. Our development set created with a subset of limited in-domain data generalizes well to the evaluation set by estimating suitable fusion parameters and by demonstrating systematic improvement. However, there is a substantial performance gap between the development and evaluation set possibly due to the large number of speakers and presence of unknown noises.

\section{Acknowledgment}
Experiments presented in this paper were partially carried out using the Grid'5000 testbed, supported by a scientific interest group hosted by Inria and including CNRS, RENATER and several Universities as well as other organizations (see \url{https://www.grid5000.fr}).

\bibliographystyle{IEEEtran}

\bibliography{mybib}


\end{document}